\documentclass[prx,reprint]{revtex4-2}
\usepackage{amsmath}
\usepackage{amssymb}
\usepackage{graphicx}
\usepackage{xcolor}

\begin{document}

\title{Neural  Monte Carlo Renormalization Group}

\author{Jui-Hui Chung}
\affiliation{Center for Theoretical Physics and Department of Physics, National Taiwan University, Taipei 10607, Taiwan}

\author{Ying-Jer Kao}
\affiliation{Center for Theoretical Physics and Department of Physics, National Taiwan University, Taipei 10607, Taiwan}
\date{\today}

\begin{abstract}
The key idea behind the renormalization group (RG) transformation is that properties of  physical systems with very different microscopic makeups can be characterized by a few universal parameters.
However, finding the optimal  RG transformation remains difficult due to the many possible choices of the weight factors in the RG procedure.
Here we show, by identifying the  conditional distribution in the restricted Boltzmann machine (RBM) and the  weight factor distribution in the RG procedure,  an optimal real-space RG transformation can be learned without prior knowledge of the physical system. 
This neural Monte Carlo RG algorithm allows for direct computation of the RG flow and critical exponents.
This scheme naturally generates a transformation that maximizes the real-space mutual information between the coarse-grained region and the environment.
Our results establish a solid connection between the RG transformation in physics and the deep architecture in machine learning, paving  the way to further interdisciplinary research. 

\end{abstract}

\maketitle

\section{Introduction}
The renormalization group (RG) \cite{wilson1971renormalization} formalism provides a systematic method for  quantitative analysis of critical phenomena. 
Among all the RG schemes, the real-space renormalization group (RSRG), first proposed by Kadanoff~\cite{kadanoff1966scaling},  is the most intuitive and natural way to perform RG transformations on lattice models~\cite{niemeijer1976renormalization}.
These methods allow for  a straightforward construction of the critical surface and  calculation of the critical exponents using numerical methods such as  Monte Carlo renormalization group (MCRG)~\cite{swendsen1979monte,wu2017variational,car2019determination}.
However, the RSRG transformation typically generates long-range couplings not present in the original Hamiltonian and truncation is necessary to make the method manageable.
From the physical point of view,  we expect the range of the \textit{renormalized} interactions of a \textit{physical} lattice system near the fixed point should not increase. 
Finding the optimal way to coarse-grain the Hamiltonian to systematically eliminate the irrelevant degrees of freedom is crucial for the success of any RSRG scheme.
The fundamental difficulty lies in the enormous degrees of freedom in choosing the weight factors for the RG transformation.
Several attempts in the past have been made to find the optimal transformation. 
Swendsen proposes an optimal MCRG scheme by introducing variational parameters into the RG procedure~\cite{swendsen1984optimization}.
Bl\"{o}te \textit{et al.} propose to modify the Hamiltonian and the weight factors such that the corrections to scaling are small~\cite{blote1996monte}.
Ron \textit{et al.} propose to choose parameters such that the critical exponent of interest was nearly constant during the MCRG iterations~\cite{ron2017surprising}.
However, it remains unclear how to determine the weight factors without  prior knowledge of the system.

The general guideline in searching for   an optimal RG transformation is to identify and eliminate the irrelevant degrees of freedom in the RG flow while retaining the relevant ones. 
However, it is difficult  \textit{a priori} to determine which degrees of freedom should be eliminated.
This resembles the question in  machine learning (ML) on how to extract relevant features from raw data.
Deep learning (DL)~\cite{lecun2015deep} using  deep neural networks (DNN) has significantly improved machine's ability in many areas such as speech recognition~\cite{Hinton:2012sp}, object recognition~\cite{Krizhevsky:2012hi},   Go and video game playing\cite{Silver:2016mw,Vinyals:2019ac,Mnih:2015jx}, as well as aided discoveries in various fields of physics~\cite{carleo2017solving,Nieuwenburg:2017cn,Carrasquilla:2017iza,Carleo:2019iw,Carrasquilla:2020er}. 
Multiple layers of representation  are used to learn distinct features directly from the training data.
The similarity between the structure of the DNN and the course-graining schemes in statistical physics inspires many efforts to establish connection  between variational RG~\cite{kadanoff1976variational} and unsupervised learning of DNN~\cite{mehta2014exact,lin2017does,schwab2016comment,koch2018mutual,Lenggenhager:2020fj,iso2018scale,funai2018thermodynamics,efthymiou2019super,chung2019optimal}.
Here, we want to address a different question: how can we  train an DNN  to obtain an  optimal RSRG transformation? 
This issue is partially addressed  from the informational theoretical perspective~\cite{koch2018mutual,Lenggenhager:2020fj}, where an optimal RG transformation is obtained by maximizing the real-space mutual information (RSMI). 
However,  the proposed RSMI algorithm requires a mutual information proxy in order to probe the effective temperature(coupling) of the system along the RG flow, rendering it less practical.
A more direct and transparent method  that  enables direct  computation of  the corresponding RG flow  and critical exponents is thus highly coveted.

 \begin{figure*}
\centering
\includegraphics[width=0.9\textwidth]{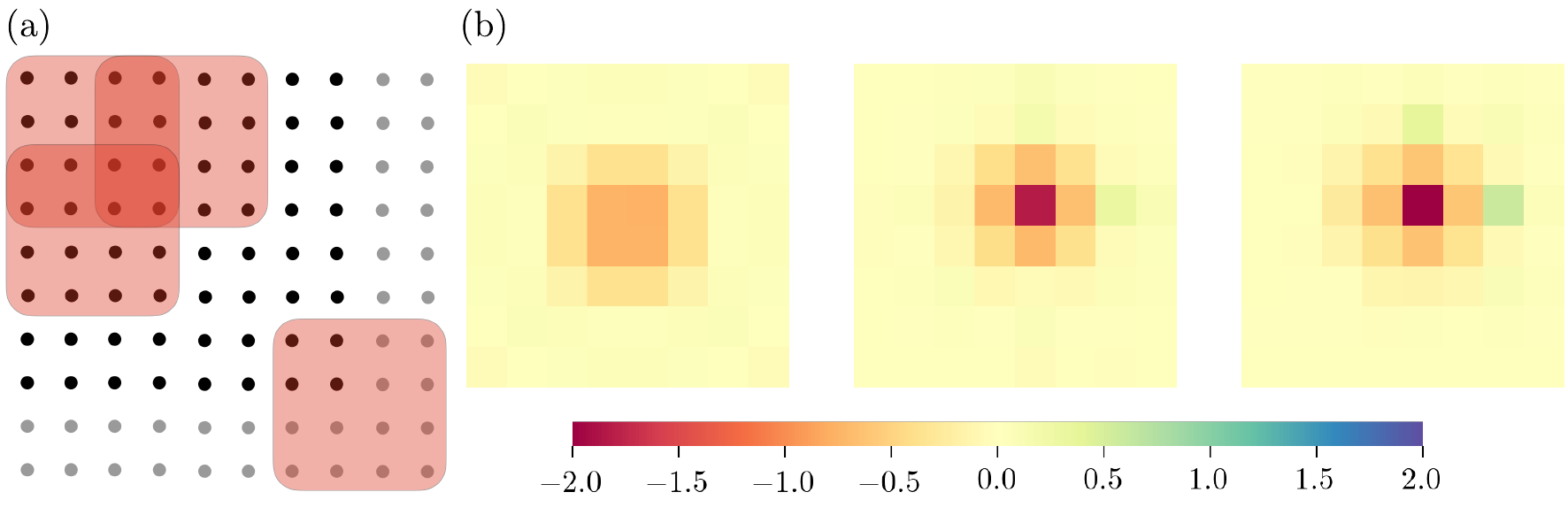}
\caption{\label{fig:filters} \textbf{RG transformation.} (a) RG transformation of  original spins (black dots) using  overlapping parametrized weight factors (red square) as in Eq.~\eqref{eq:proj}. The opaque black dots are the periodic copies of the original spins. (b) The $8\times 8$ filters are learned on a $32\times 32$ Ising model at critical NN coupling $K_1\simeq 0.4407$. From left to right, we show the development of the filters at the 10-th, 30-th and 50-th epoch corresponding to  Fig.~\ref{fig:evolution} (a).}
\end{figure*}

Here we present a scheme called neural Monte Carlo RG (NMCRG) that parametrizes the RG transformation  in terms of a restricted Boltzmann machine (RBM)~\cite{Smolensky:1986qd}.
The optimal RG transformation can  be learned by minimizing the Kullback-Leibler (KL) divergence between  the system distribution  and the marginal weight factor distribution  (defined in Eq.~\eqref{eq:WFD}).
This provides an explicit link between the RG transformation and the RBM, allowing us to use the modern ML techniques to find the optimal RG transformation. 
In addition, the scheme is readily integrated with the MCRG techniques to directly determine the effective couplings along the RG flow, and  critical exponents. 
We demonstrate the accuracy of this approach on the two-and three-dimensional classical Ising models. 
We find the optimal transformation leads to an efficient RG flow to the fixed point with short-range renormalized couplings, and  saturates the mutual information toward the upper bound.

\section{Parametrization of Real-space Renormalization Group}
Consider a generic lattice Hamiltonian,
\begin{equation}
	{H}(\sigma)=\sum_\alpha K_\alpha S_\alpha(\sigma),
\end{equation}
where the interactions $S_\alpha$ are combinations of the original spins $\sigma$ and the $K_\alpha$ are the corresponding coupling constants. 
A general RG transformation~\cite{niemeijer1976renormalization,Lenggenhager:2020fj} can be written as
\begin{equation}
	e^{{H'}(\mu)} = \sum_\sigma P(\mu|\sigma) e^{{H}(\sigma)}, \label{eq:generic_rg_transform}
\end{equation}
with  parametrized weight factors,
\begin{equation}
P(\mu |\sigma) = \frac{1}{\sum_{\mu} e^{\sum_{ij}W_{ij}\sigma_i\mu_j}}e^{\sum_{ij}W_{ij}\sigma_i\mu_j}, \label{eq:proj}
\end{equation}
where $\mu=\pm 1$ correspond to the renormalized spins in the renormalized Hamiltonian ${H}'(\mu)=\sum_\alpha K'_\alpha S_\alpha(\mu)$ with renormalized couplings $K'_\alpha$.
 $W_{ij}$ are  variational parameters to be optimized. 
In particular, if $W_{ij}$ are infinite in a local block of spins and  zero everywhere else, then we recover the majority-rule transformation~\cite{swendsen1979monte}. 
Importantly, this parameterization satisfies  the so-called trace condition 
\begin{equation}
\sum_\mu P(\mu|\sigma)=1,
\label{eq:tracecond}
\end{equation} which  is required to correctly reproduce thermodynamics~\cite{schwab2016comment,Lenggenhager:2020fj,niemeijer1976renormalization}. 
To make connection with the RBM in the following discussion, we  define the {weight factor distribution} as
\begin{equation}
P(\sigma,\mu) = \frac{1}{Z} e^{\sum_{ij}W_{ij}\sigma_i\mu_j},
\label{eq:WFD}
\end{equation}
where $Z=\sum_{\sigma,\mu}e^{\sum_{ij}W_{ij}\sigma_i\mu_j}$.
The weight factor Eq.~\eqref{eq:proj} is then simply the condition distribution of the weight factor distribution, that is, we have $P(\mu|\sigma)=P(\sigma,\mu)/\sum_\mu P(\sigma,\mu)$.

\begin{figure*}
\includegraphics[scale=0.9]{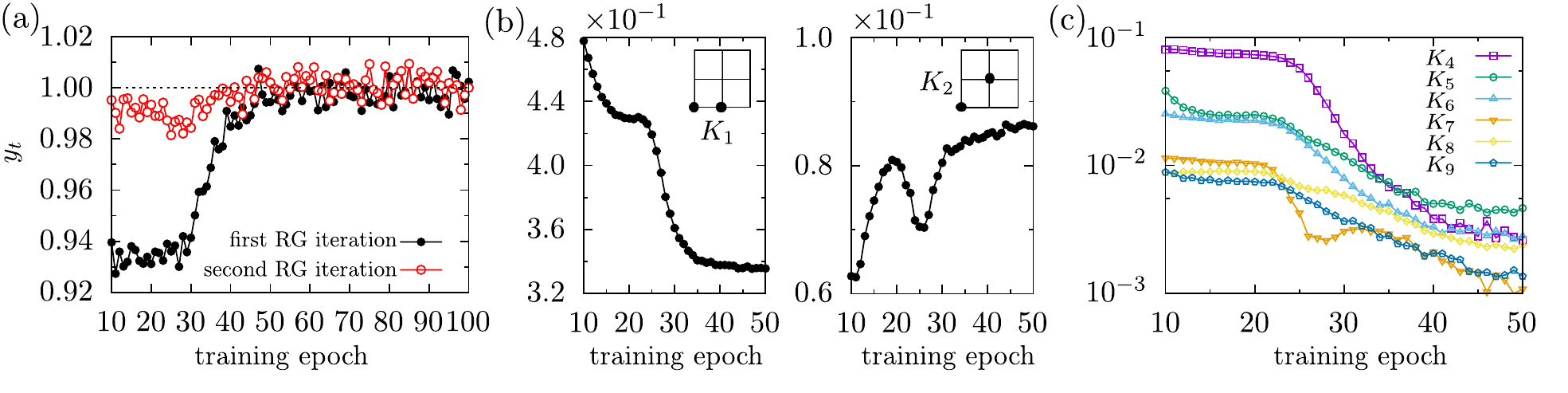}
\caption{\label{fig:evolution} \textbf{Evolution of the critical exponents and coupling parameters during training.} (a) The thermal critical exponent calculated from the weights obtained along learning process. (b) Short-range renormalized coupling parameters, nearest-neighbor $(K_1)$ and next-nearest-neighbor $(K_2)$, as a function of the training epoch. Insets indicate the corresponding couplings in real space. (c) Longer-range renormalized coupling parameters (See appendix for details). }
\end{figure*}

An RBM is a generative model that is a main staple deep learning tool to solve tasks that involve unsupervised learning~\cite{bengio2007greedy,goodfellow2016deep}.
Hidden layers of an RBM can extract meaningful features from the data~\cite{krizhevsky2009learning}.
In this regard, an RBM with fewer hidden variables than the visible variables resembles coarse-graining in RG, first pointed out by Mehta and Schwab~\cite{mehta2014exact}.
However,  their proposed mapping from the variational RG procedure to unsupervised training of a DNN does not satisfy the trace condition Eq.~\eqref{eq:tracecond},
and thus does not constitute a proper RG (See appendix for a detailed comparison). 
Here we propose a direct mapping between the RBM and the weight factors such that Eq.~\eqref{eq:tracecond} is naturally satisfied. 

An RBM can be written in terms of weights $W_{ij}$, hidden variables $h_j$ and visible variables $v_i$ as 
\begin{equation}
    Q(v,h) = \frac{1}{Z_\textrm{RBM}} e^{\sum_{ij}W_{ij}v_i h_j}, \label{eq:rbm},
\end{equation}
where $Z_\textrm{RBM}=\sum_{v,h}e^{\sum_{ij}W_{ij}v_i h_j}$.
The empirical \textit{ feature} distribution $\hat{p}'(h)$ can be extracted from the empirical distribution $\hat{p}(v)$ through
\begin{equation}
    \hat{p}'(h) = \sum_{v} Q(h|v)\hat{p}(v), \label{eq:RBM_extract_empirical}
\end{equation}
where $Q(h|v)= Q(v,h)/\sum_h Q(v,h)$ is the conditional distribution of the hidden variables, given the values of the visible variables~\cite{bengio2007greedy}. 
The optimal parameters for the RBM are chosen by minimizing the KL divergence between the empirical distribution $\hat{p}(v)$ and the marginal distribution $\sum_{h}Q(v,h)$,
\begin{equation}
   D_{\textrm{KL}}\left(\hat{p}(v) \middle\| \sum_{h}Q(v,h) \right),
\end{equation}
where  $D(p\| q) = \sum_{\sigma} p(\sigma) \log (p(\sigma)/q(\sigma))$ for two discrete distribution $p(\sigma)$ and $q(\sigma)$. 

Motivated by the similarity between Eqs.~\eqref{eq:generic_rg_transform} and \eqref{eq:RBM_extract_empirical}, we identify the conditional distribution $Q(h|v)$ in the RBM with our parametrized weight factor ${P}(\mu|\sigma)$ and associate the hidden and visible variables in the RBM with the renormalized and original spins, respectively. 
In analogy to the optimization scheme of an RBM, we propose an optimal choice of the parameters in the weight factors by minimize the KL divergence between the system distribution and the marginal weight factor distribution
\begin{equation}
D_{\textrm{KL}}\left(\frac{1}{Z}e^{{H}(\sigma)}  \middle\| \sum_{\mu} P(\sigma,\mu) \right), 
\label{eq:kl_norm}
\end{equation}
which can be carried out using standard ML techniques.

\section{Stochastic Optimization for the Optimal Criterion} \label{app:learning_contrastive}
The optimization problem is solved by the stochastic gradient descent, where the parameters are updated through decrementing them in the direction of the gradient of the KL divergence. 
We replace the system distribution $e^{H(\sigma)}/Z$ by its empirical distribution $\hat{p}(\sigma)$ over Monte Carlo samples drawn from the Wolff algorithm~\cite{Wolff:1989de} and write the KL divergence~Eq.~(\ref{eq:kl_norm}) as an expectation value over the empirical distribution 
\begin{equation}
D_{\textrm{KL}}\left(\hat{p}(\sigma) \middle\| \sum_{\mu} P(\sigma,\mu) \right). \label{eq:kl_empirical}
\end{equation}
The gradient $G_{ij}$ of the KL divergence Eq.~(\ref{eq:kl_empirical}) with respect to $W_{ij}$ can be derived as
\begin{equation}
    G_{ij}=\sum_\sigma \hat{p}(\sigma) \partial_{W_{ij}} F(\sigma) - \sum_\sigma P(\sigma) \partial_{W_{ij}} F(\sigma) ,  \label{eq:kl_gradient}
\end{equation}
where $F(\sigma)$ is the free energy defined as	$F(\sigma) = \log\sum_{\mu} e^{\sum_{ij}W_{ij}\sigma_i \mu_j}.$
The first term in Eq.~\eqref{eq:kl_gradient} is simply a sample average of the derivative of the free energy and can be readily computed.
The second term is approximated using the {contrastive divergence} algorithm \cite{hinton2002training} (CD$_k$) where the expectation value is calculated from samples drawn from a Markov chain initialized with data distribution and implemented by Gibbs sampling with $k$ Markov steps.

We  update the weights in the direction of negative gradients
\begin{equation}
W_{ij}^{(k+1)} = W^{(k)}_{ij} -  G_{ij}^{(k)},
\end{equation} 
where the superscript of the weight $W^{(k)}$ indicates the number of training epochs we have descended the weight. 
We  initialize $W^{(0)}$ randomly around zero.
Along the gradient descent we obtain a sequence of weight factors, which  can be used to compute critical exponents and  renormalized couplings, to see what feature distribution ($\hat{p}'(h)$ in Eq.\eqref{eq:RBM_extract_empirical}) the RBM is trying to \textit{learn}.
For translational-invariant systems, translational invariant parametrization of the weight factor distribution Eq.~\eqref{eq:WFD} can be achieved via convolution~\cite{lee2009convolutional}.

\begin{figure*}
\includegraphics[scale=0.92]{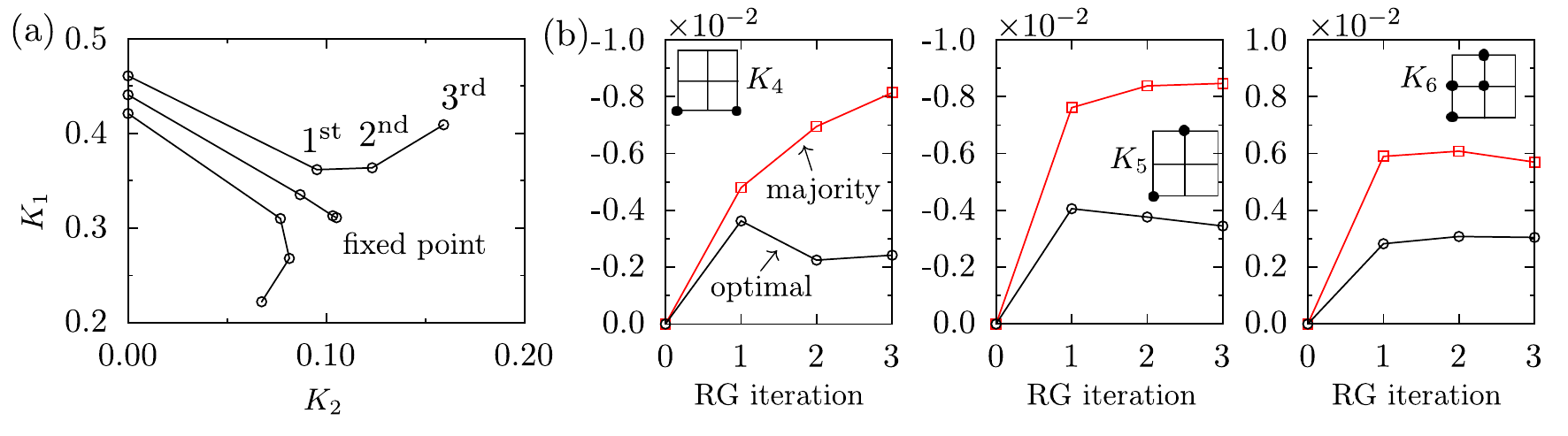}
\caption{\label{fig:traj}\textbf{Renormalization group flow.} (a) Flow of nearest- $K_1$ and next-nearest-neighbor $K_2$ coupling parameters calculated from the optimal weights along the renormalized group flow. The trajectory starts at critical couplings ($K_1\simeq 0.4407$ and $K_2=0$) and flows to the renormalized couplings at the first, second, and third RG steps. (b) Flow of long-range coupling parameters along the renormalized group trajectory for majority-rule transformation and optimal-weight transformation. Insets indicate the corresponding couplings in real space.}
\end{figure*}

\section{Two-dimensional  Ising Model }

To validate our scheme, we first consider the two-dimensional (2D) Ising model,
\begin{equation}
H(\sigma) = K_{1} S_{\text{nn}} = K_{1}\sum_{\langle{ij}\rangle} \sigma_i \sigma_j,
\end{equation}
where $\sigma_i=\pm1$,  $K_{1}$ is the nearest-neighbor coupling and $S_\textrm{nn}$ denotes the collection of  nearest-neighbor inter-spin interactions. 
In the following, we consider a 2D lattice of size $32\times 32$ with the periodic boundary condition. 
We analyze the optimal weight factors' ability to remove long-range interactions by directly calculating the renormalized couplings and extract critical exponents~\cite{swendsen1984monte}.
The computational cost of finding the optimal representation takes seconds to several minutes with a single GPU computer.

Figure~\ref{fig:filters} shows the weight factors along the optimization process (at 10th, 30th and 50th epochs corresponding to Fig.~\ref{fig:evolution} (a)) learned with a translational invariant filter of size  $8\times 8$. 
The filters are initialized uniformly around zero.
Localized features emerge after a few epochs of training and progressively aggregate toward the center,  in agreement with the conventional wisdom that renormalized and original spins  close to one another should couple more strongly than those further apart~\cite{hilhorst1978differential}. 
On the other hand, the RBM also picks up non-local correlations between the renormalized  and original spins, where the interaction strength falls off exponentially with distance.

We proceed to investigate the effect of the criterion of minimizing KL divergence to see what the machine is trying to {learn}.
In Fig.~\ref{fig:evolution} (a), we show the thermal critical exponents calculated from weight factors $W^{(k)}$ along the optimization flow. 
At the beginning of the training, the partially-optimized weight gives a poor estimate of the  thermal critical exponent at the first step of RG transformation.
After  the 30th epoch, the value grows rapidly and converges to the exact value.
In Fig.~\ref{fig:evolution} (b,c), we use the weights obtained at each training epoch to calculate the renormalized coupling parameters along the training trajectory.
The renormalized couplings, in machine-learning terms, completely describe the energy model underlying the empirical feature distribution (see Eq.~(\ref{eq:RBM_extract_empirical})) extracted by the machine for the Ising empirical distribution.
In Fig.~\ref{fig:evolution} (b), we see that the interactions are dominated by nearest ($K_1$) and next-nearest ($K_2$) neighbor couplings.
The values for the longer-range interactions flow progressively towards zero as shown in Fig.~\ref{fig:evolution} (c).
The trend shows that our optimal criterion aims to remove longer-range coupling parameters in the renormalized Hamiltonian.

 Figure~\ref{fig:traj} (a) shows the RG flow diagram projected on the short-range coupling parameters subspace for the optimal weight factors. 
The RG trajectory starting from the nearest-neighbor critical point flows rapidly to a fixed point. 
Slightly away from  the critical point, the coupling parameters flow away to the infinite (zero) temperature trivial fixed points.
Figure.~\ref{fig:traj}(b) and (c),  show the renormalized coupling parameters along the RG flow.
The coupling parameters coarse-grained with the optimal weight factors reached $K_1=0.3109(3), K_2=0.1051(2)$ and $K_3=-0.0184(2)$ at the third RG step. 
The values for longer-range interactions  are much  suppressed  compared to those obtained by the majority-rule transformation.
Since  the renormalized Hamiltonians should be dominated by short-range couplings, our \textit{learned} weight factors are superior than those for the majority-rule transformation.

\begin{table}
\caption{\label{tab:expo} \textbf{Thermal and magnetic critical exponents of the 2D Ising model} Results are obtained on a $32\times 32$ lattice using the learned optimal weight factors and the majority-rule transformation. $N_r$ is the number of RG iterations. Seven (four) coupling terms are used for even (odd) interactions. The exact values are $y_t=1$ and $y_h=1.875$.}
\begin{ruledtabular}
\begin{tabular}{c|c|l|llll}
& & & \multicolumn{4}{c}{filter size} \\
&$N_r$  & majority & 2 & 4 & 8 & 16 \\[0.5ex]
 \hline
$y_t$ &1  & 0.975(3) & 0.974(1) & 0.975(2) & 1.000(2) & 1.000(2) \\
&2  & 1.000(3) & 1.000(1) & 1.000(3) & 1.000(1) & 1.000(2) \\[0.5ex]
\hline
 $y_h$ &1  & 1.8804(2) & 1.8845(1) & 1.8887(2) &  1.8941(5) & 1.8917(1) \\
&2 & 1.8758(3) & 1.8771(1) & 1.8801(1) & 1.8827(3) & 1.8810(2)  \\[0.5ex]
\end{tabular}
\end{ruledtabular}
\end{table}

Table~\ref{tab:expo} shows the critical exponents of the 2D Ising model computed using both the RBM and majority-rule transformations.
Surprisingly, although the weights are \textit{learned}  without any prior knowledge of the model,  the exponent is very close to the exact value  at the first step of renormalization transformation giving $y_t=1.000(2)$, consistent with the exact value within the statistical error.
Equally surprising is that the RBM trained on such small training data  with only $10^4$ samples can \textit{generalize} well. 
In contrast, the majority-rule transformation  gives $y_t=0.975(3)$ at the first RG iteration.
Even though the convergence for the thermal critical exponents looks extremely good, the scheme overestimates the magnetic critical exponents in the first RG step. 
The discrepancy in the magnetic exponents is also noted previously~\cite{swendsen1984optimization,gupta1987open}.

The weight factors considered in the literature are mostly short-range \cite{kadanoff1975numerical} (decimation and majority transformation), i.e., they only couple one renormalized spin to a few original spins in the immediate vicinity. 
However, despite the seeming locality, these weight factors generally lead to an infinite proliferation of interactions upon renormalizing. 
With our proposed criterion, the learned weight factors  contain non-local terms that work as counter terms, making the renormalization transformation more local; therefore, only a few short-range interactions are produced during the RG transformation.
We note that the strategy along this line of transferring the complexity in renormalized Hamiltonian to the weight factors has yielded the first exactly soluble RG transformation \cite{hilhorst1978differential}.

\section{Three-dimensional Ising Model}
The scheme can be easily generalized to higher dimensions as long as we can train an RBM to represent the optimal RG transformation. 
Table~\ref{tab:3D} shows the thermal critical exponents computed using optimal filters starting at a system size of $64\times 64\times 64$.
The trailing numbers in the parentheses indicate the linear size of the filters.
The filters at the first ($64\to 32$) and second ($32\to 16$) steps are learned.  
The filters in the following RG steps ($16\to 8$ and $8\to 4$) use the same filter obtained in the  second step. 
We compare the results with the values obtained from the majority rule~\cite{baillie1992monte}.
Only the first twenty couplings out of the total 53 couplings in Ref.~\cite{baillie1992monte} are used. 
The $2\times 2\times 2$ optimal filter gives the exponent closest to the best estimate from the Monte Carlo  $y_t= 1.587$~\cite{Hasenbusch:2010la}. 
The $2\times 2\times 2$ optimal filter is quite homogeneous, with an average value of $0.5254(2)$, which is very close to the optimal choice 0.4314 in Ref.~\cite{ron2017surprising}.
The weight values at the second, third and forth steps are 0.5057(9), 0.510(1) and 0.544(2) respectively.

\begin{table}[tbp]
\caption{\textbf{Thermal and magnetic critical exponents of the 3D Ising model.}   Results are obtained on a $64\times 64\times 64$ lattice using the learned optimal weight factors and the majority-rule transformation. $N_r$ is the number of RG iterations. The first twenty coupling terms from \cite{baillie1992monte} are used for even and odd interactions. The accepted values  are $y_t\simeq 1.587$ and $y_h\simeq 2.482$~\cite{Hasenbusch:2010la}. \label{tab:3D}}
\begin{ruledtabular}
\begin{tabular}{c|c|l|lll}
& & & \multicolumn{3}{c}{filter size} \\
&$N_r$ & majority \cite{baillie1992monte} & 2 &  4 & 8  \\[0.5ex] 
\hline
$y_t$ &1 & 1.425(3) &  1.531(6) & 1.300(4)  & 1.323(3)\\
&2           & 1.509(2) &  1.568(2) & 1.521(2) & 1.558(2)\\
&3           & 1.547(2) &  1.579(2) & 1.556(4) & 1.568(2) \\
&4           & 1.563(9) &  1.587(3) & 1.558(6) & 1.551(3) \\
\hline
$y_h$ & 1 & 2.4578(5)  & 2.515(1)   & 2.377(1)   & 2.3819(5) \\
&2             & 2.4603(2)  & 2.4940(2) & 2.4670(2) & 2.4916(1)\\
&3             & 2.4721(4)  & 2.4875(3) & 2.4770(2) & 2.4854(3)\\
&4             & 2.476(1)    & 2.4850(8) & 2.4815(1) & 2.4845(8)\\
\end{tabular}
\end{ruledtabular}
\end{table}

\section{Real-Space Mutual Information}
We have now established that by parametrizing the weight factors as an RBM, we  can learn the optimal RG transformation. 
On the other hand, the  RSMI scheme argues that an optimal RG transformation can be obtained by maximizing the RSMI~\cite{koch2018mutual,Lenggenhager:2020fj}.
A natural question is how these two schemes are related. 
In particular, we would like to see if our optimal RG transformation also maximizes the RSMI. 

The RSMI measures the \textit{information} that the knowledge of environment degrees of freedom $\mathcal{E}$  gives about the relevant degrees of freedom $\mathcal{H}$, and is defined as
\begin{equation}
{I}(\mathcal{H};\mathcal{E}) = \sum_{\mathcal{H},\mathcal{E}} P(\mathcal{H},\mathcal{E}) \log\left(\frac{P(\mathcal{H},\mathcal{E})}{P(\mathcal{H})P(\mathcal{E})}\right)
\end{equation}
If $\mathcal{E}$ completely determines $\mathcal{H}$, then the information gained is maximized and the ${I}(\mathcal{H};\mathcal{E})$ reduces to the self-information (the entropy) of the relevant degrees of freedom $\mathcal{H}$, which itself is upper bounded by the logarithm of all possible configurations of $\mathcal{H}$.

\begin{figure}[t]
\centering
\includegraphics[width=\columnwidth]{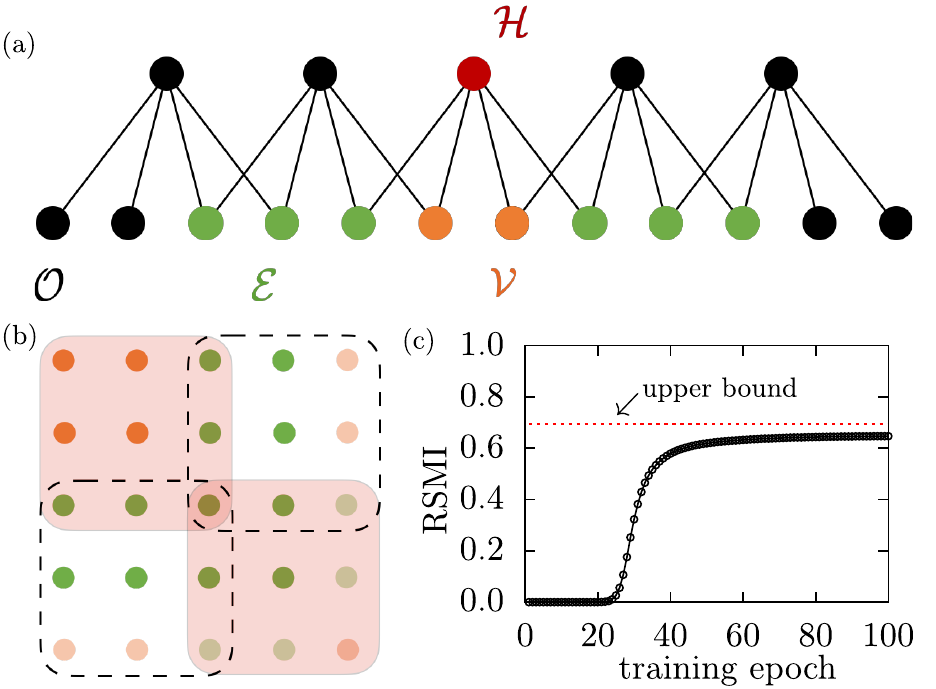}
\caption{\label{fig:small} \textbf{Real-space Mutual Information.} (a) Schematic decomposition of a system described by  a quadripartite distribution $P(\mathcal{V},\mathcal{E},\mathcal{H},\mathcal{O})$ over visible, environment, hidden and other random variables. (b) The $3\times 3$  squares represent the hidden variables which  connect to the overlapping visible variables. The opaque  dots are the \textit{periodic copies} of the visible variables. The system is partitioned into visible (orange), environment (green), hidden (top-left red square) and other (top-right, bottom-left and bottom-right red squares) random variables.  (c) As the training progresses, RSMI saturates to the upper bound $\ln 2\simeq 0.693$. }
\end{figure}

Adopting the definition in Refs.~\cite{koch2018mutual,Lenggenhager:2020fj}, we consider a system described by a quadripartite distribution $P(\mathcal{V},\mathcal{E},\mathcal{H},\mathcal{O})$ (Fig.~\ref{fig:small}(a)).
 We define the RSMI of the system as
${I}(\mathcal{H};\mathcal{E})$, i.e., 
the mutual information between hidden and environment random variables.
The relevant distributions needed to compute ${I}(\mathcal{H};\mathcal{E})$ are appropriate marginals of $P(\mathcal{V},\mathcal{E},\mathcal{H},\mathcal{O})$. 

Here we consider a $4\times 4$ Ising model with the periodic boundary condition where RSMI can be  can computed exactly. 
We train a $3\times 3$ filter on the system to obtain an optimal weight factor distribution. 
Figure~\ref{fig:small}(a) and (b) shows the partition of the lattice into visible (orange), environment (green), hidden (top-left red square) and other (top-right, bottom-left and bottom-right red squares) random variables. 
Figure~\ref{fig:small} (c) shows the evolution of RSMI during training. 
Random initialization of the filters gives a zero RSMI, and as the training progresses, the RSMI saturates to the upper bound  $\ln 2\simeq 0.693$. 
This shows clearly that the optimal weight factors obtained from our algorithm saturate the RSMI as proposed in Refs.~\cite{koch2018mutual,Lenggenhager:2020fj}.
However, our scheme allows for a direct calculation of the renormalized coupling parameters and critical exponents using the MCRG algorithms without resorting to proxy systems. 

\section{Conclusions}
We demonstrate a scheme based on RBM that is capable of learning the optimal RG transformation from Monte Carlo samples. 
The similarity between the standard RBM and the weight factors means that we can take advantage of the progress in the ML architectures and techniques  to parameterize and train the filters for RG. 
This algorithm is flexible and can  be easily applied to disordered systems~\cite{wang1988monte}. 
Although we focus on the RBM with binary variables, for models with continuous variables such as XY or Heisenberg models, one can use Gaussian-Bernoulli RBMs to better model the RG transformation~\cite{Hinton:2006rt}. 
Generalization of the current scheme to quantum systems should be straightforward by the quantum-to-classical mapping of the $d$-dimensional quantum system to $d+1$-dimensional classical system~\cite{novotny1985monte}. 
It would be interesting to test the NMCRG scheme on fermionic systems to see how fermionic sign manifests itself. 
Finally, we note that in the 2D Ising model, the filter  has to reach the size of $8\times 8$  to obtain reasonable critical exponents, while in the  3D case, a $2\times 2\times 2$ filter suffices to give the best result. 
Wether this can be associated with the logarithmic correction in the 2D Ising model warrants further studies~\cite{Barma:1984jx}.

\begin{acknowledgments}
 This work was supported by Ministry of Science and Technology (MOST) of Taiwan under Grants No. 108-2112-M-002-020-MY3 and No. 107-2112-M-002-016-MY3, and partly supported by National Center of Theoretical Science (NCTS) of Taiwan. We are grateful to the National Center for High-performance Computing for computer time and facilities. The code that generates data used in this paper is available at \url{https://github.com/unixtomato/nmcrg}.

\end{acknowledgments}

\appendix


\section{Monte Carlo Renormalization Group}\label{app:MCRG}
Here we summarize the MCRG method used to calculate the critical exponents and renormalized coupling parameters from Monte Carlo samples for a given filter~ \cite{swendsen1984monte}.
To determine the critical exponents, we need to calculate the derivatives of transformation
\begin{equation}
	T^{(n+1)}_{\alpha\beta}\equiv \frac{\partial K^{(n+1)}_\alpha}{\partial K^{(n)}_\beta},
\end{equation}
which is given by the solution of the linear equation \cite{swendsen1979monte}
\begin{equation}
	\frac{\partial\langle{S^{(n+1)}_\gamma}\rangle}{\partial K^{(n)}_\beta} = \sum_\alpha \frac{\partial\langle{S^{(n+1)}_\gamma}\rangle}{\partial K^{(n+1)}_\alpha}\frac{\partial K^{(n+1)}_\alpha}{\partial K^{(n)}_\beta}.
\end{equation}
Here $\langle{S_\gamma^{(n)}}\rangle$ is the expectation of the spin combinations at the $n$th RG iterations. 
The derivatives of these expectation value of the spin combinations are obtained from the correlation functions 
\begin{eqnarray}
	&&\frac{\partial\langle{S^{(n+1)}_\gamma}\rangle}{\partial K^{(n)}_\beta} = \langle{S^{(n+1)}_\gamma S^{(n)}_\beta}\rangle - \langle{S^{(n+1)}_\gamma}\rangle\langle{S^{(n)}_\beta}\rangle,\\
	&&\frac{\partial\langle{S^{(n+1)}_\gamma}\rangle}{\partial K^{(n+1)}_\alpha} = \langle{S^{(n+1)}_\gamma S^{(n+1)}_\alpha}\rangle \nonumber \\
	&&\quad\quad\quad\quad\quad\quad\quad\quad - \langle{S^{(n+1)}_\gamma}\rangle\langle{S^{(n+1)}_\alpha}\rangle.
\end{eqnarray}

Given a set of spin configurations sampled from some Hamiltonian ${H}=\sum_\alpha K_\alpha S_\alpha$, we would like to infer back the coupling parameters of ${H}$.
Define a specific spin-dependent expectation
\begin{equation}
\langle S_{\alpha,l}\rangle_l \equiv \frac{1}{z_l} \sum_{\sigma_l} S_{\alpha,l} e^{\mathcal{H}_l},
\end{equation}
where
$z_l=\sum_{\sigma_l}e^{{H}_l}$ and ${H}_l = \sum_\alpha K_\alpha S_{\alpha,l}$ and $S_{\alpha,l}$ are combination of spins in $S_{\alpha}$ that includes only $\sigma_l$. 
Here $z_l$ and ${H}_l$ and hence $\langle S_{\alpha,l}\rangle_l$ depend on spins neighboring to $\sigma_l$. 
The summation of $\sigma_l$ can be carried out analytically and we obtain the formula
\begin{equation}
\langle S_{\alpha,l}\rangle_l =  \widehat{S}_{\alpha,l} \tanh\left[\sum_\beta K_\beta \widehat{S}_{\beta,l}\right],
\end{equation}
where $S_{\alpha,l}\equiv \sigma_l\widehat{S}_{\alpha,l}$. 

The correlation functions can then be written in another form as
\begin{equation}
\frac{1}{Z}\sum_{\sigma}  S_{\alpha} e^{{H}} = \frac{1}{Z}\sum_{\sigma} \left[\frac{1}{m_\alpha}\sum_l \langle S_{\alpha,l}\rangle_l\right] e^{H(\sigma)},
\end{equation}
where $m_\alpha$ is the number of spins in the combination $S_\alpha$.
Introducing a second set of coupling parameters $\{\widetilde{K}_\alpha\}$ we define
\begin{equation}
\langle\widetilde{S}_\alpha\rangle = \frac{1}{Z}\sum_{\sigma} \left\{\frac{1}{m_\alpha}\sum_l \widehat{S}_{\alpha,l} \tanh\left[\sum_\beta \widetilde{K}_\beta \widehat{S}_{\beta,l}\right] \right\} e^{H(\sigma)}.
\end{equation}
It can be shown that $\{\langle S_\alpha\rangle\}=\{\langle\widetilde{S}_\alpha\rangle\}$ if and only if $\{K_\alpha\}=\{\widetilde{K}_\alpha\}$.

\begin{figure}
\includegraphics[scale=1]{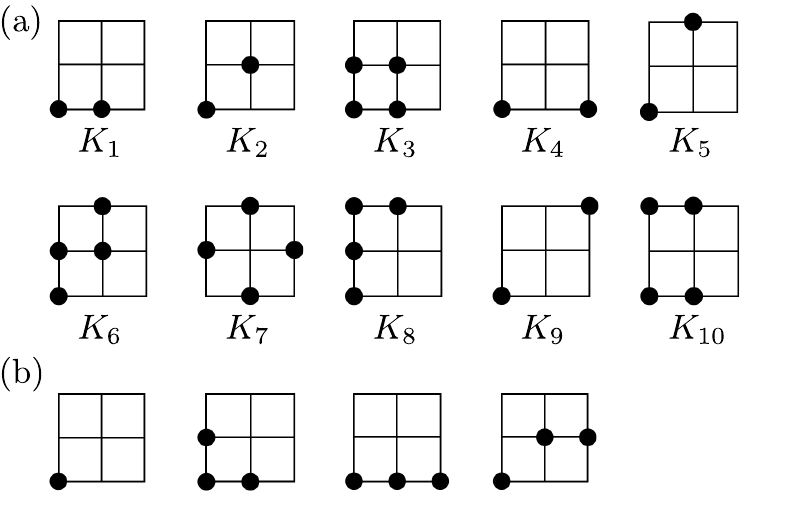}
\caption{\label{fig:operators} \textbf{2D couplings.} (a) Couplings used for the calculation of renormalized coupling parameters. The first seven are used for the calculation of the thermal critical exponent. (b) The four couplings used to compute the magnetic critical exponent.}
\end{figure}

Figure~\ref{fig:operators} shows the couplings used for the calculation of the renormalized coupling parameters for the two-dimensional Ising model. 
The first seven even couplings in (a) are used to compute the thermal critical exponent.  
The odd couplings in (b) are used to compute the magnetic critical exponent. 

\section{Comparison with Other RBM-based Schemes}
\subsection{RG transformation and Normalizing Condition}
Consider again a general RG transformation
\begin{equation}
e^{H'(\mu)} = \sum_\sigma P(\mu|\sigma) e^{H(\sigma)}, \label{rg}
\end{equation}
where $P(\mu|\sigma)$ is  the weight factor. 
The weight factor is required to satisfy the \textit{trace condition} 
\begin{equation}
\sum_\mu P(\mu|\sigma) = 1.
\label{eq:tr}
\end{equation}
We argue that the trace condition is indispensable, since the condition leads to the invariance of free energy under renormalization and the following fundamental relation
\begin{equation}
f(K) = b^{-d} f(K'), \label{fundamental}
\end{equation}
where $f(K)$ is the free energy density of the system in the thermodynamic limit. For $K$ consisting of nearest-neighbor coupling and magnetic field, under suitable transformation, we could arrive at $f(t,h)=b^{-d}f(b^{y_t} t,b^{y_h} h)$ where $y_t$ and $y_h$ are the often sought-after critical thermal and magnetic exponents. 

In the following, we review the schemes proposed in Refs.~\cite{mehta2014exact} and~\cite{koch2018mutual} and point out the shortcomings in each scheme. 

\subsection{Variational RG and Mehta and Schwab's Mapping}
In Ref.~\cite{mehta2014exact}, the weight factor is defined as
\begin{equation}
P_W(\mu|\sigma) = e^{\sum_{ij}W_{ij}\sigma_i\mu_j - H(\sigma)}. \label{mehta}
\end{equation}
Here $H(\sigma)$ is the original Hamiltonian, e.g., $H(\sigma) = K\sum_{\langle ij\rangle}\sigma_i\sigma_j.$ 
The $W_{ij}$'s are the {variational parameters.} 
The form of the weight factor does not satisfy the trace condition and, in general, it is not possible to choose the parameters $W_{ij}$ to satisfy the trace condition \eqref{eq:tr}. 
The fundamental relation \eqref{fundamental} is only approximated. 

We note that in the original procedure of variational renormalization group \cite{kadanoff1976variational}, the form of the weight factor is chosen with variational parameters such that for all values of variational parameters the weight factor must satisfy the trace condition. 
The variational parameters are used, instead, to optimize the lower bound of the approximated free energy density.

Define a distribution of the weight factor with variational parameters $W_{ij}$, 
\begin{equation}
P_W(\sigma) = \frac{\sum_\mu e^{\sum_{ij}W_{ij}\sigma_i\mu_j}}{\sum_\sigma\sum_\mu e^{\sum_{ij}W_{ij}\sigma_i\mu_j}}.
\end{equation}
In Ref.~\cite{mehta2014exact} the variational parameters are chosen to make 
\begin{equation}
D_{\text{KL}}\left(\frac{e^{H(\sigma)}}{Z} \middle\| P_W(\sigma) \right), \label{kullback}
\end{equation}
as small as possible. This completely fixes the variational parameters, leaving no room for optimizing the lower bound free energy approximation. 
That is to say, the \textit{variational} approximation in machine learning \eqref{kullback} and the \textit{variational} approximation of thevariational renormalization theory work at completely different levels.

The rationale of the criterion \eqref{kullback} for choosing the variational parameters is that it is a necessary but \textit{not} sufficient condition for the trace condition to be satisfied
\[ 
\sum_\mu e^{\sum_{ij}W_{ij}\sigma_i\mu_j - H(\sigma)} = 1
\]
 implies 
 \[
 e^{H(\sigma)} = \sum_\mu e^{\sum_{ij}W_{ij}\sigma_i\mu_j}.
 \]
The normalization factor $\sum_\sigma\sum_\mu e^{\sum_{ij}W_{ij}\sigma_i\mu_j}$ is  equal to the partition function for the original Hamiltonian, denoted as $Z$. 
Therefore the divergence (\ref{kullback}) is exactly zero. 
The criterion is not sufficient since when \\\[
e^{f(\sigma)}/\sum e^{f(\sigma)}=e^{g(\sigma)}/\sum e^{g(\sigma)},
\]
we have 
\[
e^{f(\sigma)-g(\sigma)}= \sum e^{f(\sigma)}/\sum e^{g(\sigma)}
\]
 where the trace condition fails up to some unknown constant not necessarily equal to one.

On the other hand, with the parametrized form of weight factor as in (\ref{mehta}), the renormalized Hamiltonian would then describe the marginal distribution $P_W(\mu)$ of the RBM. Define $P_W(\mu)$ to be
\begin{equation}
P_W(\mu) = \frac{\sum_\sigma e^{\sum_{ij}W_{ij}\sigma_i\mu_j}}{\sum_\sigma\sum_\mu e^{\sum_{ij}W_{ij}\sigma_i\mu_j}}.
\end{equation}
Carrying out the RG transformation (\ref{rg}) for the weight factors (\ref{mehta}) gives
\begin{equation}
e^{H'(\mu)} = \sum_\sigma e^{\sum_{ij}W_{ij}\sigma_i\mu_j - H(\sigma)} e^{H(\sigma)} = \sum_\sigma e^{\sum_{ij}W_{ij}\sigma_i\mu_j}.
\end{equation}
{The normalization factor $\sum_\sigma\sum_\mu e^{\sum_{ij}W_{ij}\sigma_i\mu_j}$ is thus equal to the partition function, $Z'$, for the renormalized Hamiltonian irrespective of the choice of the variational parameters $W_{ij}$. }
Therefore 
\begin{equation}
P_W(\mu)= \frac{e^{H'(\mu)}}{Z'}. 
\end{equation}
In this respect, we can say that the \textit{hidden variables of the machine} is described by the renormalized Hamiltonian.

\subsection{{Real-space Mutual Information Algorithm}}

\begin{figure}[t]
\centering
\includegraphics[width=\columnwidth]{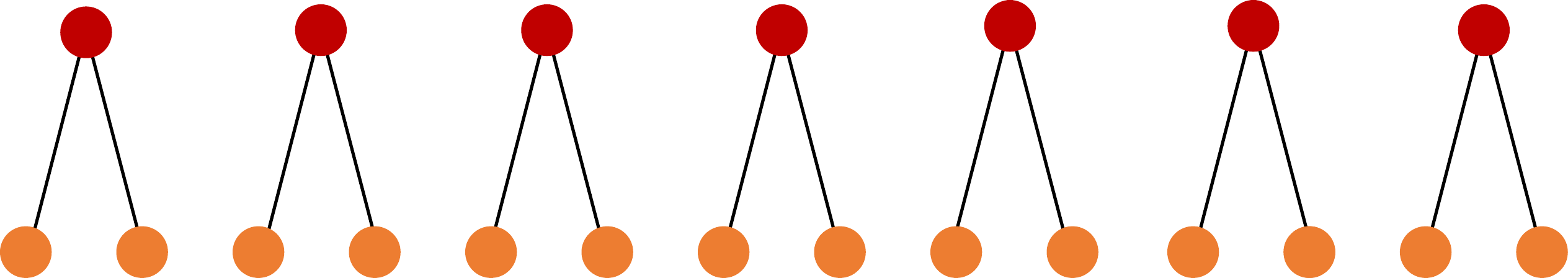}
\caption{\label{fig:rsmi_coarse}Weight factor factorized as identical copies of local weight factors.}
\end{figure}

\begin{figure}[t]
\centering
\includegraphics[width=\columnwidth]{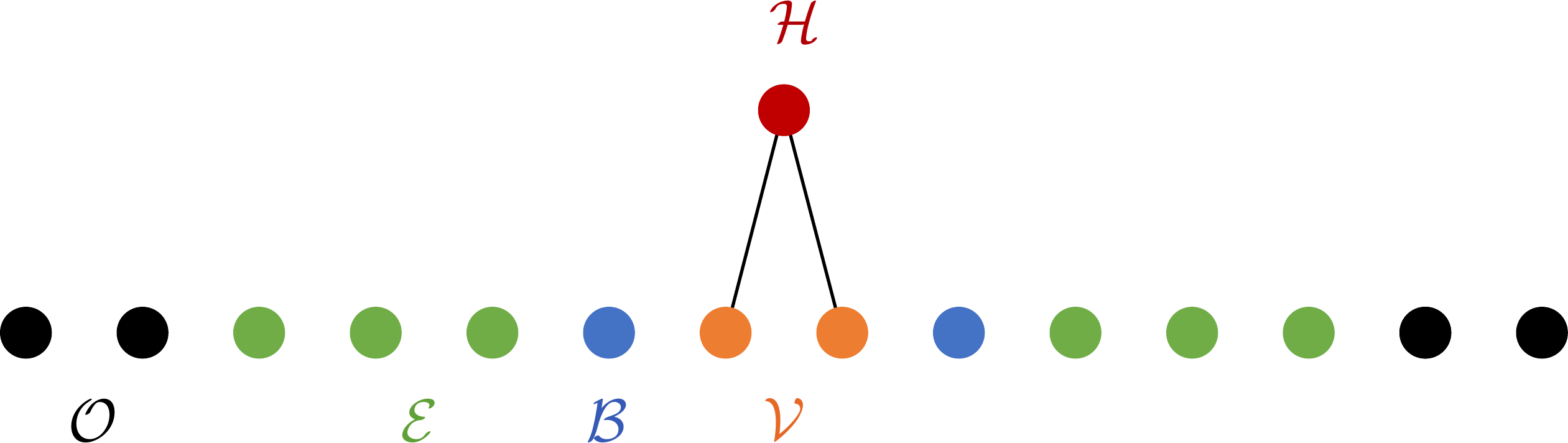}
\caption{\label{fig:rsmi_decomp}Schematic decomposition of the system spins into the visible($\mathcal{V}$), buffer($\mathcal{B}$), environment($\mathcal{E}$) and other( $\mathcal{O}$) spins respectively. 
 We refer to a local block of hidden spins $\mathcal{H}$.}
\end{figure}

In Ref.~\cite{koch2018mutual}, the weight factor factorizes as
\begin{equation}
P_\Lambda(\mu|\sigma) = \prod_j P_\Lambda(\mathcal{H}_j|\mathcal{V}_j)
\end{equation}
where $\mathcal{H}_j=\{\mu_j\}$ consists of a single renormalized spin and $\mathcal{V}_j=\{\sigma_j^1,\sigma_j^2\}$ consists of two original spins in the case of one-dimension system (and $2\times 2$ in the case of two-dimensional system), see Fig. \ref{fig:rsmi_coarse}. 
The \textit{local} weight factor is parametrized as
\begin{equation}
P_\Lambda(\mathcal{H}_j| \mathcal{V}_j) = \frac{e^{\sum_i\Lambda_i\mu_j \sigma_j^i}}{\sum_{\mu}e^{\sum_i\Lambda_i\mu_j \sigma_j^i}}.
\end{equation}
The variational parameters $\Lambda$ is obtained through only a single copy of the local weight factor and hence we omit the subscript $j$ in the following. 
Consider a single copy of the local weight factor where the local visible spins $(\mathcal{V})$ are embedded among the buffer $(\mathcal{B})$, environment $(\mathcal{E})$ and other $(\mathcal{O})$ spins which collectively form the original system spins $(\mathcal{X})$, see Fig.~\ref{fig:rsmi_decomp}. 
Construct two proxies $P_{\Theta_1}(\mathcal{V})$ and $P_{\Theta_2}(\mathcal{V},\mathcal{E})$ in the form of RBMs trained on the restriction of $\mathcal{X}=(\mathcal{V},\mathcal{B},\mathcal{E},\mathcal{O})$ MC samples (from the Boltzmann equilibrium distribution of the Hamiltonian of concerned). 
Define $P_\Lambda(\mathcal{E},\mathcal{H}) = \sum_{\mathcal{V}} P_{\Theta_2}(\mathcal{V},\mathcal{E}) P_\Lambda (\mathcal{H}|\mathcal{V})$  and $P_\Lambda(\mathcal{H})=\sum_\mathcal{V} P_{\Theta_1}(\mathcal{V}) P_\Lambda (\mathcal{H}|\mathcal{V}) $ and $P (\mathcal{E}) =\sum_{\mathcal{V}} P_{\Theta_2}(\mathcal{V},\mathcal{E}).$
The variational parameters $\Lambda$ are chosen to make
\begin{equation}
I_\Lambda(\mathcal{H};\mathcal{E}) = \sum_{\mathcal{H},\mathcal{E}} P_\Lambda(\mathcal{E},\mathcal{H}) \log\left(\frac{P_\Lambda(\mathcal{E},\mathcal{H})}{P_\Lambda(\mathcal{H})P(\mathcal{E})}\right)
\end{equation}
as large as possible, where the distributions needed in the right hand side are defined as above. 
Since $P(\mathcal{E})$ is independent of $\Lambda$, we instead maximize a proxy $A_\Lambda=\sum_{\mathcal{H},\mathcal{E}} P_\Lambda(\mathcal{E},\mathcal{H}) \log\left({P_\Lambda(\mathcal{E},\mathcal{H})}/{P_\Lambda(\mathcal{H})}\right)$ of mutual information.
However, to evaluate the proxy $A_\Lambda$, further approximations have to be made.

In order to perform quantitative analysis, the authors construct a ``thermometer''  function $T(A_\Lambda)$ which maps the proxy $A_\Lambda$ to the temperature. 
The thermometer works to extract \textit{effective temperature} of the renormalized system. 
To construct such a thermometer, it is required to generate sets of MC samples at different temperatures. 
For each set of samples, one can compute the proxy $A_\Lambda$ and hence know the mapping from $A_\Lambda$ to the temperature $T$ for this set of samples. 
For a given type of system (e.g., Ising), we can write $T(A_\Lambda)$ as $T(T_0,L,b,l)$ where $T_0$ is the temperature of the initially prepared system, $L$ is the initial system size, $b$ is the scale factor, and $l$ is the scaling length ($l=0$ means the original system and $l=1$ means one-step renormalization and so on). We can then \textit{fit} a function to these sets of samples and construct the thermometer. 
M. Koch-Janusz and Z. Ringel  postulate a {scaling function} of the form $f((L/b^l)^{1/\nu})$ related to the effective renormalized temperature $T(T_0,L,b,l)$ as
\begin{equation}
\frac{T(T_0,L,b,l)-T_c}{T_0-T_c} = f((L/b^l)^{1/\nu}),
\end{equation}
where $T_c$ is the critical temperature of the original system. 
Finally one could \textit{collapse} the plot of $(T-T_c)/(T_0-T_c)$ as a function of $(L/b^l)^{1/\nu}$ to estimate the value of $\nu$ and $T_c$.

\subsection{Neural Monte Carlo {Renormalization Group}}
In our work, we define the weight factor to be
\begin{equation}
P_W(\mu|\sigma) = \frac{e^{\sum_{ij}W_{ij}\sigma_i\mu_j}}{\sum_\mu e^{\sum_{ij}W_{ij}\sigma_i\mu_j}}. \label{optimal_weight}
\end{equation}
{where, for translational invariant system, the variational parameters are \textit{shift invariant}, that is, for different $j$ and $j'$ we have }
\begin{equation}
W_{ij}=W_{((i+j'-j)\text{mod} N)j'}
\end{equation}
in the case of one-dimensional system.
The weight factor satisfies the trace condition for all values of $W_{ij}$'s. Let us define a joint distribution out of this weight factor
\begin{equation}
P_W(\mu,\sigma) = \frac{e^{\sum_{ij}W_{ij}\sigma_i\mu_j}}{\sum_\sigma\sum_\mu e^{\sum_{ij}W_{ij}\sigma_i\mu_j}}.
\end{equation}
Here $P_W(\mu,\sigma)$ has exactly the same form of a RBM and the weight factor can be viewd as the conditional distribution $P_W(\mu|\sigma) = P_W(\mu,\sigma)/\sum_\mu P_W(\mu,\sigma)$.

Consider one of the breakthrough in the realm of deep learning where Hinton introduced a greedy layer-wise unsupervised learning algorithm (See Sec. 2.3 of \cite{bengio2007greedy}). Denote $P_W(\mu|\sigma)$ the posterior over $\mu$ associated with the \textit{trained} RBM (we recall that $\sigma$ is the observed input). This gives rise to a (feature) empirical distribution $p'(\mu)$ over the hidden variables $\mu$ when $\sigma$ is sampled from the data empirical distribution $p(\sigma)$: we have
\begin{equation}
p'(\mu) = \sum_\sigma P_W(\mu|\sigma) p(\sigma). \label{extract}
\end{equation}
The samples of $\mu$ with empirical distribution $p'(\mu)$ become the input for another layer of RBM. We can view RBM to work as extracting features $\mu$ from inputs $\sigma$.

Note the similarity between RG transformation (\ref{rg}) and the feature extraction process (\ref{extract}). We could \textit{postulate} that the input distribution $p(\sigma)$ is determined by some Hamiltonian $H(\sigma)$ where $p(\sigma)=e^{H(\sigma)}/Z.$ We postulate that the posterior distribution $P_W(\mu|\sigma)$ of an RBM works as a weight factor to do RG transformation: $e^{H'(\mu)}=\sum_\sigma P_W(\mu|\sigma)e^{H(\sigma)}.$ Hence the feature extraction process (\ref{extract}) becomes a necessary condition for the system to perform the RG transformation. 
In other words, the \textit{feature distribution extracted by the machine} is described by the renormalized Hamiltonian.

Now the variational parameters in the weight factor $P_W(\mu|\sigma)$ are free to change. All choices of parameters should derive a well-defined RG transformation. 
The criterion for choosing the parameters is entirely arbitrary from the perspective of doing RG: we do not know \textit{a priori} what weights $W_{ij}$'s could give a ``nicer'' RG flow. A nice RG flow, however, should bring the original Hamiltonian closer to the fixed point fast.
 Also, it should remove long-range coupling parameters for practical purposes of performing RG and, loosely speaking, for killing the irrelevant scaling fields.
Critical exponents and the coupling parameters can be easily computed using the MCRG techniques described in the previous section.

In the realm of machine learning, the weights of an RBM are chosen to make the divergence (\ref{kullback}) as small as possible. 
We note that the criterion is entirely machine-learning-theoretical. 
In contrast, in Ref.~\cite{mehta2014exact}, the criterion also serves as a necessary condition for the weight factor to satisfy the trace condition, a notion which is RG-theoretical.

In summary, our NMCRG scheme provides an ansatz for the weight factors in the RG transformation  such that the trace condition is always satisfied and the optimal RG transformation can be learned.
It also allows for a direct computation of the renormalized coupling parameters and critical exponents. As demonstrated in the main text, the MNCRG scheme also naturally saturates RSMI. 
The simplicity and flexibility of the scheme should find more applications in the future.

\bibliography{draft}

\end{document}